\documentclass[aps,twocolumn,showpacs,amsmath,amssymb,nofootinbib,superscriptaddress,showkeys]{revtex4-1}

\usepackage{epsfig}
\usepackage{slashed}

\makeatletter
\def\slashchar#1{{\mathpalette\c@ncel{#1}}} 
\makeatother

\begin{document}

\title{Light Flavour and Heavy Quark Spin Symmetry
in Heavy Meson Molecules}
 \author{C. Hidalgo-Duque}

 \author{J. Nieves}
 \affiliation{Instituto de F\'{\i}sica Corpuscular (IFIC), Centro
   Mixto CSIC-Universidad de Valencia, Institutos de Investigaci\'on
   de Paterna, Aptd. 22085, E-46071 Valencia, Spain}

 \author{M. Pav\'on Valderrama}\email{pavonvalderrama@ipno.in2p3.fr}
 \affiliation{Institut de Physique Nucl\'eaire, Universit\'e Paris-Sud,
    IN2P3/CNRS, F-91406 Orsay Cedex, France}

\date{\today}

\begin{abstract}
\rule{0ex}{3ex}
We propose an effective field theory incorporating light SU(3)-flavour
and heavy quark spin symmetry to describe charmed
meson-antimeson bound states.
At lowest order the effective field theory entails a remarkable simplification:
it only involves contact range interactions among the heavy meson
and antimeson fields.
We show that the isospin violating decays of the $X(3872)$ can be used
to constraint the interaction between the $D$ and a $\bar{D}^*$ mesons
in the isovector channel.
As a consequence, we can rule out the existence of a isovector partner
of the $X(3872)$.
If we additionally assume that the $X(3915)$ and $Y(4140)$ are $D^*\bar{D}^*$
and $D_s^*\bar{D}_s^*$ molecular states, we can determine the full spectrum
of molecular states with isospin $I=0$, $\frac{1}{2}$ and $1$.

\end{abstract}


\maketitle

\section{Introduction}

The $X(3872)$ resonance~\cite{Choi:2003ue} 
has opened new perspectives in hadron spectroscopy.
The $X(3872)$, even though it clearly contains a $c\bar{c}$ pair,
does not fit well within the standard charmonium spectrum.
Disentangling its nature requires a more exotic explanation.
Among the theoretical proposals available, the interpretation of the $X(3872)$
as a hadronic molecule~\cite{Voloshin:1976ap,Tornqvist:1991ks,Tornqvist:1993ng,Manohar:1992nd,Ericson:1993wy}
is especially promising and has attracted the attention of the community.
The motivation behind the molecular picture,
in which the $X(3872)$ is a bound state of a charmed meson and antimeson,
is the striking closeness of this resonance to the $\rm D^0 \bar{D}^{0*}$
threshold, $m_{X(3872)} = 3871.68 \pm 0.17 \,{\rm MeV}$,
to be compared with $m_{D^0} + m_{D^{0*}} = 3871.84 \pm 0.20\,{\rm MeV}$
~\cite{Beringer:1900zz}.
However, the likelihood of the molecular hypothesis depends on
the $J^{PC}$ quantum numbers of the $X(3872)$,
which have not been experimentally determined yet.
They are either $1^{++}$ or $2^{-+}$~\cite{Abe:2005iya,Abulencia:2006ma,delAmoSanchez:2010jr} (see also the interesting analysis of Ref.~\cite{Hanhart:2011tn}),
of which only $1^{++}$ is compatible with a low-lying S-wave bound state.

The discovery of the $X(3872)$ has been followed by the experimental observation
of a series of hidden charm resonances above the open-charm threshold,
the so-called XYZ states.
Some of the XYZ states fit well within the charmonium spectrum
(most notably the $Z(3940)$~\cite{Uehara:2005qd}), but 
others do not and may require, just like the $X(3872)$,
non-conventional explanations.
A few might be molecular:
several authors~\cite{Liu:2009ei,Branz:2009yt,Ding:2009vd}
have proposed the $X(3915)$~\cite{Uehara:2009tx}
and $Y(4140)$~\cite{Aaltonen:2009tz}
to be $D^*\bar{D}^*$ and $D_s^*\bar{D}_s^*$ bound states.
The $Y(4260)$~\cite{Aubert:2005rm} might even have a three body structure
($J/\Psi\,K\bar{K}$)~\cite{MartinezTorres:2009xb}.
The $Y(4660)$~\cite{Wang:2007ea} and $X(4630)$~\cite{Pakhlova:2008vn}
have been theorized to be a $f_0(980) \Psi'$
molecule~\cite{Guo:2008zg,Guo:2010tk}
(they may be the same state~\cite{Guo:2010tk}).
More recently, the Belle collaboration has observed two hidden bottom
resonances, the $Z_b(10610)$ and $Z_b(10650)$~\cite{Collaboration:2011gj,Belle:2011aa}, located just a few MeV away from the $\rm B\bar{B}^*$
and $\rm B^*\bar{B}^*$ thresholds.
Hence, the $Z_b$'s may also have
a molecular nature~\cite{Voloshin:2011qa,Cleven:2011gp,Mehen:2011yh}.

Heavy meson-antimeson molecules are very interesting objects
from the theoretical point of view.
As far as they are not tightly bound, the meson and antimeson will preserve
their individuality and will not probe the specific details of the
short range interaction responsible of their  binding.
Thus, there exists a scale separation and molecular states may be amenable 
to an effective field theory (EFT) treatment in which they can be described as
mesons interacting via contact interactions and pion exchanges
(e.g. X-EFT~\cite{Fleming:2007rp}, Heavy Meson Molecule EFT~\cite{Valderrama:2012jv}).
In addition, the heavy-light quark content of heavy meson molecules
implies a high degree of symmetry.
They are subjected to heavy quark spin symmetry (HQSS)~\cite{Isgur:1989vq,Isgur:1989ed,Neubert:1993mb,Manohar:2000dt}, which imposes
interesting constraints in the heavy meson-antimeson
interaction~\cite{AlFiky:2005jd}.
As a consequence, HQSS can be used to predict the existence of
so far unobserved molecular states~\cite{Voloshin:2011qa,Mehen:2011yh,Nieves:2012tt}.
On the other hand, we must not forget the light quark content of heavy meson molecules.
If we consider $q = u,d,s$, we expect SU(3) flavour symmetry to hold:
we can arrange molecular states within SU(3) multiplets.
In this work we will explore the consequences of this symmetry in the spectrum
of heavy meson molecules.

The effect of HQSS and SU(3) flavour symmetry is to generate relationships
among the heavy meson-antimesons interactions in different channels.
Four parameters are enough to describe the twenty-four possible S-wave
molecules\footnote{This is the total dimension, without considering the
  spin-isospin third component multiplicities of the $H\bar H$ space
  ($H= D^+,D^0,D_s^0, D^{*+},D^{*0}$ and $D_s^{*0}$) or states that
  are connected by a C-parity transformation, such as $D_s \bar{D}$
  and $\bar{D}_s D$ that are not counted twice.}.
That is, we need four data points to predict the full molecular spectrum.
For this purpose we will assume the molecular nature of certain XYZ states
such as the $X(3872)$, $X(3915)$ and $Y(4140)$.
If the predictions turn out to be correct, this will serve
as a confirmation that the states used as input are molecular.
We notice that the $X(3915)$ could also be a charmonium state,
as the decay properties are similar to a molecule's,
with the most important decay channels
being $J / \Psi \, \omega$ and $D\bar{D}$.
However, according to Ref.~\cite{Guo:2012tv}, the charmonium hypothesis
implies a $D\bar{D}$ partial decay width of about one hundred ${\rm MeV}$
at least, larger than the experimental total decay width
(about $30\,{\rm MeV}$).
In contrast, the molecular picture predicts a partial decay width
to $D\bar{D}$ of the order of tens of ${\rm MeV}$~\cite{Nieves:2012tt}.
We also warn that the $Y(4140)$ is far from being confirmed and
might not even exist: after its discovery~\cite{Aaltonen:2009tz},
it has not been observed in subsequent
experiments~\cite{Shen:2009vs,Aaij:2012pz}.
Finally, we notice that SU(2) isospin and SU(3) flavour symmetry are broken
in the masses of the heavy mesons.
From this we can derive interesting consequences.
In particular, from  the analysis of the isospin violating branching 
of the $X(3872)$ into $J / \Psi\,\omega$ and $J / \Psi\,\rho$,
we will be able to determine the strength of the contact interaction
in the isovector channel and to discard the existence of an
isovector or $D_s \bar{D}_s^*$ counterpart
of the $X(3872)$

The article is structured as follows: in Sect.~\ref{sec:EFT-LO} we briefly
review the effective field theory formalism that we use for the description
of heavy meson molecules.
In Sect.~\ref{sec:isobreaking} we consider the isospin violating decay
of the $X(3872)$ into $J / \Psi \, \rho$, which can be used to obtain
information about the interaction of the $D\bar{D}^*$
in the isovector channel.
In Sect.~\ref{sec:partners}
we calculate the location of the SU(3)-flavour and HQSS
partners of the $X(3872)$, $X(3915)$ and $Y(4140)$.
Finally, we present our conclusions in Sect.~\ref{sec:conclusion}.

\section{The EFT Description at Lowest Order}
\label{sec:EFT-LO}

In this section we review the EFT we use for describing heavy meson molecules.
The presentation is brief -- we have already discussed
this EFT in previous publications~\cite{Valderrama:2012jv,Nieves:2012tt} -- 
and we concentrate on the ideas rather than the technicalities.
The EFT description involves pions and heavy meson/antimeson fields
and the local interactions among these degrees of freedom that
are compatible with the known low energy symmetries,
most notably HQSS and chiral symmetry.
A remarkable simplification is that pion exchanges are weaker
than naively expected~\cite{Valderrama:2012jv} and only
enter as a perturbation at subleading orders.
A similar thing happens to coupled channel dynamics.
Hence, at lowest or leading order (${\rm LO}$),
the EFT consists on heavy meson and antimeson interacting
through a contact range potential.

\subsection{Overview of the EFT Formalism}

EFTs are generic theoretical descriptions of low energy phenomena.
They become an adequate  tool  for situations in which a more
fundamental description in terms of the underlying high-energy dynamics
is impractical for whatever reasons.
Thus they are very useful for low-energy hadronic processes
where quantum chromodynamics (QCD) is not solvable
owing to asymptotic freedom and confinement.

The formulation of an EFT requires the identification of the degrees of
freedom and symmetries that are important for the low energy dynamics.
For heavy meson molecules, the degrees of freedom are the heavy meson
and antimeson, plus the pion field.
The relevant symmetries, which provide the connection to the underlying
theory (QCD), are chiral symmetry and HQSS. 
It is important to notice that EFTs are only useful if there exists
a separation of scales between low and high energy physics.
If we name the characteristic low energy scale at which the EFT is expected
to work as $Q$, and the high energy scale as $\Lambda_0$,
EFTs allows to construct the amplitudes as a power series expansion
in terms of the small parameter $x_0 = Q / \Lambda_0$
\begin{eqnarray}
\mathcal{A} = \sum_{\nu} {\mathcal{A}}^{(\nu)} =
\sum_{\nu} x_0^{\nu}\, \hat{\mathcal{A}}^{(\nu)} \, ,
\end{eqnarray}
meaning that if we truncate the expansion at $\nu = \nu_{\rm max}$,
the {\it a priori} error of the calculation
will be $x_0^{\nu_{\rm max} + 1}$.
The ordering principle behind the EFT expansion is power counting.
It is related to the scaling properties of the diagrams contributing
to the EFT amplitude $\mathcal{A}$ under a rescaling
of all the $Q$'s
\begin{eqnarray}
\mathcal{A}^{(D)} (\lambda\,Q, \Lambda_0) = \lambda^{\nu_D}\,
\mathcal{A}^{(D)} (Q, \Lambda_0) \, ,
\end{eqnarray}
where the superscript $(D)$ indicates that we are considering a single
diagram of order $\nu_D$ (equivalently, we say that $D$ is of order
$Q^{\nu_D}$).
At each order $Q^{\nu}$ in the EFT expansion there is only a finite
number of diagrams involving the low energy fields and
their symmetries.

\subsection{The EFT Potential}

Heavy mesons are non-relativistic and form bound states.
The first property entails a simplification of the EFT description,
while the second triggers a series of non-trivial changes
in the power counting that we will comment later.
The simplification is that the heavy meson-antimeson potential
is a well-defined quantity.
Instead of directly expanding the scattering amplitude within EFT,
we can expand (and truncate) the heavy meson-antimeson potential
\begin{eqnarray}
V = \sum_{\nu = 0}^{\nu_{\rm max}} V^{(\nu)}
+ \mathcal{O}(x_0^{\nu_{\rm max} + 1}) \, .
\end{eqnarray}
Then we calculate wave functions and observables in the standard
quantum mechanical fashion.
For example, we can generate bound states by iterating the EFT potential
in the Schr\"odinger / Lippmann-Schwinger equation~\footnote{
This is in fact the basis of the Weinberg counting
as originally formulated in the context of two-nucleon scattering
in EFT~\cite{Weinberg:1990rz,Weinberg:1991um}.
However, the power counting we obtain is equivalent to the Kaplan, Savage
and Wise proposal~\cite{Kaplan:1998tg,Kaplan:1998we}, modulo the obvious
difference that here we are dealing with heavy mesons
instead of nucleons.
}.

At ${\rm LO}$, which corresponds to $\nu = 0$,
the heavy meson-antimeson potential is local
\begin{eqnarray}
\langle \vec{p}\,' | V^{(0)} | \vec{p} \rangle =
V^{(0)}(\vec{p}\,' - \vec{p}\,) \, ,
\end{eqnarray}
and receives the contribution of two diagrams, a four heavy meson vertex
and the one pion exchange (OPE) potential.
The ${\rm LO}$ potential reads (schematically)
\begin{eqnarray}
\label{eq:V-EFT-lowest-order}
V^{(0)}(\vec{q}\,) &=& C_0^{(0)} + \eta \frac{g^2}{2
  f_{\pi}^2}\, \frac{(\vec{a} \cdot \vec{q}\,)\,(\vec{b} \cdot
  \vec{q}\,)}{\vec{q}^{\,\,2} + m_{\pi}^2} \, .
\end{eqnarray}
The contact range coupling $C_0^{(0)}$ is a free parameter.
Its value can be determined from the location of a bound state.
The OPE contribution depends on the pion decay constant
$f_{\pi} \simeq 132\,{\rm MeV}$ and the axial coupling $g \simeq 0.6$,
which we have particularized for the charmed meson case.
The sign $\eta$ and the spin operators $\vec{a}$ and $\vec{b}$ depend on 
whether the incoming/outgoing heavy meson and antimeson are
pseudoscalar or vector (the details can be consulted
in Ref.~\cite{Valderrama:2012jv}).

A problematic feature of the EFT potential is that it tends to a constant,
non-zero value at large exchanged momentum.
Thus, heavy meson-antimeson loops containing the EFT potential will
be divergent.
We solve this issue by renormalizing the EFT calculations.
For that, we define a regularized potential $V_{\Lambda}$ as follows
\begin{eqnarray}
\langle \vec{p}\,' | V_{\Lambda} | \vec{p}\, \rangle = 
f(\frac{p'}{\Lambda})\,
\langle \vec{p}\,' | V | \vec{p}\, \rangle \,
f(\frac{p}{\Lambda}) \, ,
\end{eqnarray}
where $\Lambda$ is an ultraviolet cut-off, $p=|\vec{p}\,|$,
$p'=|\vec{p}\,'|$   and $f(x)$ a regulator function
obeying the conditions (i) $f(0) = 1$, (ii) $f(x) \to 0$ for $x \to \infty$.
For concreteness we use the gaussian regulator function
\begin{equation}
f_{\rm Gauss}(x) = e^{-x^2} \, ,
\end{equation}
though other choices of the regulator are equally valid~\footnote{
We notice that multipolar regulators are closer in spirit
to the form factors that are usually employed
in more phenomenological approaches.
Yet in EFT treatments there is no privileged choice of a regulator function.
Nonetheless we will briefly comment on the impact of using
other regulators later on when we present our results.
}.
The cut-off dependence of the theory is absorbed in the counterterms,
completing the renormalization process.
We comment that there is no universal criterion for choosing the cut-off,
but as a general rule it does not need to be much larger
than the high energy scale $\Lambda_0$~\cite{Lepage:1997cs}.
Here we employ the window $\Lambda = 0.5-1.0\,{\rm GeV}$,
which is of the order of $\Lambda_0$.

\subsection{Power Counting and Bound States}

The existence of bound states implies changes in the power counting
of the potential.
This can be easily appreciated by considering the bound state equation
\begin{eqnarray}
| \Psi_B \rangle = G_0(E) V | \Psi_B \rangle \, ,
\end{eqnarray}
where $| \Psi_B \rangle$ is the wave function, $G_0(E) = 1 / (E - H_0)$
the resolvent operator and $V$ the non-relativistic potential.
Then it is apparent from the EFT point of view that power counting
requires the successive iterations of $G_0 V$ to be of the same order
\begin{eqnarray}
\mathcal{O}(G_0 V) = \mathcal{O}(G_0 V G_0 V) \, .
\end{eqnarray}
If we take into account that $G_0(E)$ is of order $Q$, it is clear
that the existence of a bound states requires that the EFT potential
contains a contribution of order $Q^{-1}$.
That is, the naive assignment of order $Q^0$ to the lowest order
potential is incompatible with the existence of bound states.
The solution is to promote at least one of the diagrams conforming
the $Q^0$ potential to order $Q^{-1}$~\cite{Kaplan:1998tg,Kaplan:1998we,Gegelia:1998gn,Birse:1998dk,vanKolck:1998bw}.
For heavy meson-antimeson molecules we usually move the contact range potential
to $Q^{-1}$ and left the OPE potential at $Q^0$.
The consequences of this promotion are that (i) we redefine ${\rm LO}$ as
$Q^{-1}$ and (ii) OPE is a subleading order effect, that is,
a small perturbation over the ${\rm LO}$ results.
As we are only considering the ${\rm LO}$ calculation, we are left
with a very simple theory that only contains contact interactions.
The exception is the isoscalar bottom sector, in which OPE is stronger than
naively expected and is moved to order $Q^{-1}$ together
with the counterterm~\cite{Valderrama:2012jv}.

To end the discussion, we find it useful to write down the eigenvalue
equation for the bound states:
\begin{eqnarray}
\frac{1}{C_0(\Lambda)} = - \int \, \frac{d^3 \vec{q}}{(2\pi)^3} \,
f^2(\frac{q}{\Lambda})\,\frac{2\mu}{q^2 + \gamma^2} \, ,
\end{eqnarray}
which relates the position of the bound state with the strength of the
contact interaction $C_0$ (i.e. we are assuming that this is
the only operator that enters at ${\rm LO}$).
In the equation above $f(x)$ is the regulator function and
$\gamma^2 = -2\mu\,E_B$, with $E_B < 0$ the bound
state energy and $\mu$ the reduced mass of
the two-body system.
Provided the cut-off $\Lambda$ and the wave number $\gamma$,
we can determine the value of $C_0$.
Conversely, from $C_0$ and $\Lambda$, we can predict
the location of a bound state.
Further details on the eigenvalue equation above and on the wave functions
we obtain from it can be found in Refs.~\cite{Nieves:2012tt,Nieves:2011zz}.

\subsection{The Heavy Quark Spin Structure of the EFT Potential}

According to the previous arguments, at lowest order in the EFT expansion
the potential only contains an energy- and momentum- independent S-wave
contact range interaction (per channel).
In a given isospin--strangeness sector, naively this translates into six
counterterms, one for each of the possible S-wave configurations of a
heavy meson molecule.
However, HQSS reduces the number of independent counterterms to two,
which we call $C_a$ and $C_b$~\cite{AlFiky:2005jd}.

Owing to HQSS the ${\rm LO}$ potential mixes the different particle channels.
The reason is that the $P$ ($\bar{P}$) and $P^*$ ($\bar{P}^*$) mesons
(antimesons) can be transformed into each other by means of a flip of
the spin of the heavy quark.
Nevertheless the total angular momentum and parity of a heavy meson molecule
is conserved.
If we consider the following set of basis (with well-defined $J^{P}$)
\begin{eqnarray}
\mathcal{B}(0^{+}) &=&
\left\{ | P\bar{P} \rangle , | P^*\bar{P}^* (0) \rangle \right\} \, , \\
\mathcal{B}(1^{+}) &=&
\left\{ | P\bar{P}^* \rangle, | P^*\bar{P} \rangle ,
| P^*\bar{P}^*(1) \rangle \right\} \, , \\
\mathcal{B}(2^{+}) &=&
\left\{ | P^*\bar{P}^*(2) \rangle \right\} \, ,
\end{eqnarray}
the ${\rm LO}$ potential takes the form
\begin{eqnarray}
V^{\rm LO}(\vec{q}, {0^{+}}) &=& 
\begin{pmatrix}
C_{a} & \sqrt{3}\,C_{b} \label{eq:contact1}\\
\sqrt{3}\,C_{b} & C_{a} - 2\,C_{b}
\end{pmatrix} \, , \\
V^{\rm LO} (\vec{q}, 1^{+}) &=&
\begin{pmatrix}
\phantom{+}C_{a} & -C_{b} & \sqrt{2}\,C_{b} \\
-C_{b} & \phantom{+}C_{a} & \sqrt{2}\,C_{b} \\
\sqrt{2}\,C_{b} & \sqrt{2}\,C_{b} & C_{a} - C_{b} \\
\end{pmatrix} \, , \\
V^{\rm LO}(\vec{q}, 2^{+}) &=& C_{a} + C_{b} \label{eq:contact2-b} \, .
\end{eqnarray}
The notation $| P^{(*)} \bar{P}^{(*)} \rangle$ is used to indicate a
system of a heavy meson/antimeson containing a heavy quark/antiquark,
regardless of the light quark content.
That is, the heavy meson $P^{(*)}$ is not necessarily the antiparticle of
the heavy antimeson $\bar{P}^{(*)}$, as they could contain different
types of light quarks.
The number in parenthesis in the $|P^*\bar{P}^*(J) \rangle$ states 
is the total intrinsic spin $J$ to which the two vector meson
system couples.

In the specific case in with the heavy meson and antimeson are each other
antiparticle, the C-parity is a well-defined (and conserved)
quantum number.
The $J^P = 0^{+}$ and $2^{+}$ states are now $J^{PC} = 0^{++}$ and $2^{++}$
states, but neither the potentials nor the bases we use to express them
change
\begin{eqnarray}
\mathcal{B}(0^{++}) &=& \mathcal{B}(0^{+}) \, , \\
\mathcal{B}(2^{++}) &=& \mathcal{B}(2^{+}) \, , \\
\nonumber \\
V^{\rm LO}(\vec{q}, {0^{++}}) &=& V^{\rm LO}(\vec{q}, {0^{+}}) \, , \\
V^{\rm LO}(\vec{q}, {2^{++}}) &=& V^{\rm LO}(\vec{q}, {2^{+}}) \, .
\end{eqnarray}
In contrast, the $1^+$ states can be further subdivided into 
$1^{++}$ and $1^{+-}$ states that do not mix under the
effect of the ${\rm LO}$ potential.
We can write a $1^{++}$ and $1^{+-}$ basis
\begin{eqnarray}
\mathcal{B}(1^{+-}) &=&
\left\{ \frac{1}{\sqrt{2}}\left( | P\bar{P}^* \rangle + | P^*\bar{P} \rangle
\right) , | P^*\bar{P}^*(1) \rangle \right\} \, , \\
\mathcal{B}(1^{++}) &=&
\left\{ \frac{1}{\sqrt{2}}\left( | P\bar{P}^* \rangle - | P^*\bar{P} \rangle
\right) \right\} \, , 
\end{eqnarray}
for which the ${\rm LO}$ potential now reads
\begin{eqnarray}
V^{\rm LO} (\vec{q}, 1^{+-}) &=&
\begin{pmatrix}
C_{a} - C_{b} & 2\,C_{b} \\
2\,C_{b} & C_{a} - C_{b}
\end{pmatrix} \, , \\
V^{\rm LO}(\vec{q}, 1^{++}) &=& C_{a} + C_{b} \label{eq:contact2-a} \, .
\end{eqnarray} 
As can be seen, the $1^{++}$ heavy meson-antimeson molecule
decouples from the two $1^{+-}$ components.
In addition, the $1^{++}$ potential is identical to the $2^{++}$ one.
A $1^{++}$ heavy meson-antimeson molecule implies a $2^{++}$ HQSS
partner~\cite{Nieves:2012tt}.

There is still a significant simplification in the EFT potential
that we have not discussed yet.
In the charm sector the EFT description of heavy meson molecules is
only expected to be valid if the binding energy is smaller than
a maximum value $B_{\rm max}$ of the order of
$150-300\,{\rm MeV}$
(that is, we take $B_{\rm max} = \gamma_{\rm max}^2 / 2 \mu$,
with $\gamma_{\rm max} = \Lambda_0 \sim 0.5-1 \,{\rm GeV}$ and $\mu = m_D / 2$,
where $m_D$ is the $D$ meson mass).
This figure is similar to the mass gap among the heavy meson-antimeson
particle channels: 
\begin{eqnarray}
\Delta(0^{+}) \equiv M(D^*\bar{D}^*) - M(D\bar{D}) &\simeq& 280\,{\rm MeV} \, , \\
\Delta(1^{+}) \equiv M(D^*\bar{D}^*) - M(D\bar{D}^*) &\simeq& 140\,{\rm MeV} \, , 
\end{eqnarray}
where $\Delta(J^P)$ denotes the gap in the $J^P$ channel.
The prize of ignoring the couped channel effects is a reduction of
the expansion parameter of the theory from $x_0 = \sqrt{{B}/{B_{\rm max}}}$
to $x_0' = \sqrt{{B}/{\Delta(J^P)}}$, where $B$ is the binding energy of
the molecular state and we have included the square roots to
translate the energy ratios into momentum ratios
(we are using a non-relativistic EFT).
As the sizes of $B_{\rm max}$ and $\Delta(J^P)$ are comparable,
we can safely ignore the coupled channel dynamics without
compromising the range of validity of the EFT description
of the bound states. 
Moreover, explicit calculations carried out in Ref.~\cite{Nieves:2012tt,Nieves:2011zz} for the isoscalar channels confirm that the size of
the coupled channel effects is the expected one in EFT.

Thus, we can simplify the EFT potential in the $0^{+}$ and $1^{+}$
channels to
\begin{eqnarray}
V^{\rm LO}_{{\rm P \bar P}}(\vec{q}, {0^{+}}) &=& C_{a} \label{eq:coni} \, , \\
V^{\rm LO}_{{\rm P^* {\bar P}^* }}(\vec{q}, {0^{+}})
&=& C_{a} - 2\,C_{b} \, , \\
V^{\rm LO}_{{\rm P^* {\bar P} / P {\bar P}^*}}(\vec{q}, {1^{+}})
&=&
\begin{pmatrix}
\phantom{+}C_{a} & -C_{b} \\
-C_{b} & \phantom{+}C_{a}
\end{pmatrix}
\, , \\ 
V^{\rm LO}_{{\rm P^* {\bar P}^* }}(\vec{q}, {1^{+}})
&=& C_{a} - C_{b} \, , \label{eq:conf} 
\end{eqnarray}
where we only need to keep track of the coupled channel dynamics
in the $1^{+}$ $P^* {\bar P} / P {\bar P}^*$ case.
If C-parity is a good quantum number, the $P^* {\bar P}$ and  $P {\bar P}^*$
thresholds coincide and the $1^{+-}$ component separates from the
$1^{++}$ one:
\begin{eqnarray}
V^{\rm LO}_{{\rm P^* {\bar P} / P {\bar P}^*}}(\vec{q}, {1^{+-}})
&=& C_{a} - C_{b} \\
V^{\rm LO}_{{\rm P^* {\bar P} / P {\bar P}^*}}(\vec{q}, {1^{++}})
&=& C_{a} + C_{b}
\, ,
\end{eqnarray}
while there is no change in the other channels (except the additional
C-parity quantum number) 
\begin{eqnarray}
V^{\rm LO}_{{\rm P \bar P}}(\vec{q}, {0^{++}}) &=&
V^{\rm LO}_{{\rm P \bar P}}(\vec{q}, {0^{+}}) \, , \\
V^{\rm LO}_{{\rm P^* {\bar P}^* }}(\vec{q}, {0^{++}})
&=& V^{\rm LO}_{{\rm P^* {\bar P}^* }}(\vec{q}, {0^{+}}) \, , \\
V^{\rm LO}_{{\rm P^* {\bar P}^* }}(\vec{q}, {1^{+-}})
&=& V^{\rm LO}_{{\rm P^* {\bar P}^* }}(\vec{q}, {1^{+}}) \, .
\end{eqnarray}
Now the potentials in the $1^{+-}$  $P^* {\bar P} / P {\bar P}^*$ 
and $P^* \bar{P}^*$ channels coincide.
That is, the $1^{+-}$ molecular states come in pairs,
one state per each of the particle channels,
as happens with the $Z_b(10610)$ and $Z_b(10650)$ in the bottom sector.

\subsection{The SU(3)-Flavour Structure of the EFT Potential}

SU(3)-flavour symmetry implies that we can organize
the heavy meson-antimeson states into SU(3)-multiplets.
As there is a light quark and antiquark ($3$ and $\bar{3}$ representations),
heavy meson-antimeson molecules can be organized in a singlet
and octet representation, $3 \otimes \bar{3} = 1 \oplus 8$.
We can label the heavy meson with a flavour index, that is, we write
$P_a$, $\bar{P}^a$,  $P^*_a$ and $\bar{P}^{*a}$ with $a = 1,2,3$,
where the quark content is
\begin{eqnarray}
\label{eq:quark-content}
P_a, P_a^* = \left (Q \bar{u}, Q \bar{d}, Q \bar{s} \right )
~ \mbox{and} ~
\bar{P}^a, \bar{P}^{*a} =
\begin{pmatrix}
\bar{Q} u \\
\bar{Q} d \\
\bar{Q} s
\end{pmatrix} \, ,
\end{eqnarray}
and $Q = c, b$ represents the heavy quark.
In this notation we can construct the singlet and octet representations
as follows
\begin{eqnarray}
| P\bar{P}, 1\rangle &=& \frac{1}{\sqrt{3}}\,|P_a\bar{P}^a\rangle \, , \\
| P\bar{P}, 8; j\rangle &=&
\frac{1}{\sqrt{2}}\,(\lambda_j)^a_{\,b}|P_a\bar{P}^b\rangle
\, , 
\end{eqnarray}
plus the analogous expressions for $P\bar{P}^*$, $P^*\bar{P}$
and $P^*\bar{P}^*$, where the $\lambda_{j}$'s are the Gell-Mann matrices.
SU(3) flavour symmetry implies that the heavy meson-antimeson interaction
distinguishes the singlet and octet representations
\begin{eqnarray}
\langle P\bar{P}, 1 | V^{\rm LO} | P\bar{P}, 1 \rangle &=& C^{(1)}_a \, , \\
\langle P\bar{P}, 8; i | V^{\rm LO} | P\bar{P}, 8; j
\rangle &=& C^{(8)}_a \, \delta_{ij} \, , 
\end{eqnarray}
and do not mix them
\begin{eqnarray}
\langle P\bar{P}, 1 | V^{\rm LO} | P\bar{P}, 8; j \rangle &=& 0 \, , 
\end{eqnarray}
where we have particularized for the case of a $P\bar{P}$ molecule
with quantum numbers $J^{PC} = 0^{++}$ and ignored the particle coupled
channel dynamics stemming from HQSS.
The extension to other particle channels and $J^{PC}$ quantum numbers
is trivial and only entails the substitution of $C^{(\mu)}_a$
in the equation above by the adequate linear combination
of $C^{(\mu)}_a$ and $C^{(\mu)}_b$, with $\mu =1$ or $8$.

Flavour symmetry also conserves the strangeness $S$ and
the isospin $I$ of the heavy meson-antimeson system.
For the singlet representation we have $S=0$ and $I=0$.
For the octet, however, we can have different values of the quantum numbers
$S$ and $I$, and thus we may find more convenient to write them in the basis
\begin{eqnarray}
| P\bar{P}, 8; S; I, M_I\rangle
\end{eqnarray}
instead of $| P\bar{P}, 8; j\rangle$,
where the relation between the two basis can be readily obtained
(the only technicality involved is the isospin phase convention
for the light antiquarks).
In this basis we have
\begin{eqnarray}
\langle P\bar{P}, 8; S; I; M_I | V^{\rm LO} | P\bar{P}, 8; S'; I'; M_I'
\rangle \nonumber \\
\qquad \qquad \qquad
= C^{(8)}_a \, \delta_{SS'} \, \delta_{II'} \, \delta_{M_I M_I'} \, , 
\end{eqnarray}
plus the corresponding expressions for the particle channels ($P\bar
P^*, P^*\bar{P}$ and $P^*\bar{P}^*$). The conservation of $S$, $I$ and $M_I$ is now evident.

However, flavour symmetry is broken at the level of particle masses.
In general, this only represents a problem for isoscalar states
with hidden strangeness ($I=0$ and $S=0$), where the threshold
of the particle channel that contains the $s\bar{s}$
light quark pair is considerably above the $u\bar{u}$
and $d\bar{d}$ channels.
For example, in the charm sector the energy gap between the $D_s\bar{D}_s$
and the $D^0\bar{D}^0$/$D^+\bar{D}^-$ thresholds is about $200\,{\rm MeV}$,
of the order of the maximum binding energy we expect to be able
to describe within the EFT framework.
In other cases, the particle channels within a given $S$ and $I$
channel have approximately the same masses.
Isospin symmetry breaking is of the order of a few MeV and we will not take
it into account unless the molecular state we are describing is very shallow,
as in the $X(3872)$, a case we will consider in detail
at the end of this section.

For the treatment of SU(3) breaking we define
\begin{eqnarray}
| P\bar{P}\,[0] \rangle &=& \frac{1}{\sqrt{2}}\,\left(
| P_1\bar{P}^1 \rangle + |P_2\bar{P}^2 \rangle  \right) \, , \\
| P_s\bar{P}_s \rangle &=& | P_3\bar{P}^3 \rangle \, ,
\end{eqnarray}
that is, we distinguish between the SU(2) isoscalar ($I=0$) state
and the hidden strangeness state, and the indices $1$, $2$ and $3$
refer to the representation of Eq.~(\ref{eq:quark-content}).
In this basis we can rewrite the singlet and octet isoscalar representations as 
\begin{eqnarray}
|{P\bar{P}}, 1\rangle &=& \frac{1}{\sqrt{3}}
\left[ \sqrt{2}\,| {P\bar{P}}[0] \rangle + 
| {P_s\bar{P}_s} \rangle  \right] \, , \\
|{P\bar{P}}, 8;0;0,0\rangle &=& \frac{1}{\sqrt{3}}
\left[ | {P\bar{P}}[0] \rangle - 
\sqrt{2} | {P_s\bar{P}_s} \rangle \right] \, .
\end{eqnarray}
By inverting these relations, we find that in the basis
\begin{eqnarray}
\mathcal{B} = \left\{ | P\bar{P}\,[0] \rangle, | P_s\bar{P}_s \rangle \right\}
\, ,
\end{eqnarray}
the flavour symmetric interaction can be written as
\begin{eqnarray}
V^{\rm LO} =
\begin{pmatrix}
\frac{2}{3} C^{(1)}_a + \frac{1}{3} C^{(8)}_a & \frac{\sqrt{2}}{3}\left [
C^{(1)}_a - C^{(8)}_a \right] \\ 
\frac{\sqrt{2}}{3}\left [ C^{(1)}_a - C^{(8)}_a\right] & \frac{1}{3} C^{(1)}_a + \frac{2}{3} C^{(8)}_a  
\end{pmatrix} \, ,
\end{eqnarray}
where flavour symmetry is broken owing to the energy gap
among the $|P\bar{P}[0] \rangle$ and $| P_s\bar{P}_s \rangle$ thresholds.
As commented in the previous paragraph, the energy gap is large enough
as to justify the treatment of the isoscalar and hidden strangeness
channels as uncoupled.

In terms of notation we find that working in the SU(2) isospin basis
is easier than in the SU(3) one.
If we define the SU(2) isovector state (with $M_I = 0$) as
\begin{eqnarray}
| P\bar{P}[1] \rangle &=& \frac{1}{\sqrt{2}}\,\left(
|P_1\bar{P}^1 \rangle - |P_2\bar{P}^2 \rangle  \right) \, ,
\end{eqnarray}
we can rewrite the EFT potential in the SU(2) isospin basis as
\begin{eqnarray}
\langle P\bar{P}[0] | V^{\rm LO} | P\bar{P}[0] \rangle &=& C_{0a} \, \label{eq:c0a}, \\
\langle P\bar{P}[1] | V^{\rm LO} | P\bar{P}[1] \rangle &=& C_{1a} \, \label{eq:c1a} ,
\end{eqnarray}
plus the corresponding expressions for the other particle ($P\bar
P^*, P^*\bar{P}$ and $P^*\bar{P}^*$) / spin / C-parity
combinations, which require the inclusion of the new counterterms
$C_{0b}$ and $C_{1b}$.

The counterterms $C_{0a}$ and $C_{1a}$ introduced in
Eqs.~(\ref{eq:c0a}) and (\ref{eq:c1a}) are related to the ones
in the original SU(3)-multiplet basis by
\begin{eqnarray}
C_{0a} &=& \frac{2}{3} C^{(1)}_a + \frac{1}{3} C^{(8)}_a , \\
C_{1a} &=& C^{(8)}_a \, .
\end{eqnarray}
In the SU(2) basis, the interaction in the $| P_s\bar{P_s} \rangle$
channel reads
\begin{eqnarray}
\langle P_s\bar{P}_s| V^{\rm LO} | P_s\bar{P}_s \rangle &=& 
\frac{1}{2}\,\left( C_{0a} + C_{1a} \right) \, ,
\end{eqnarray}
that is, the average of the isoscalar and isovector contact range potentials.

Finally, we consider the $X(3872)$, where isospin breaking is important
as this bound state is especially shallow.
The energy gap between the $D^0\bar{D}^{*0}$ and $D^+ {D}^{*-}$
is $8\,{\rm MeV}$, which is smaller than the binding energy of
the $X(3872)$ (about $4\,{\rm MeV}$ in the isospin
symmetric limit, or, equivalently, almost at
the $D^0\bar{D}^{*0}$ threshold).
What we do then is to treat the neutral ($D^0\bar{D}^{*0}$) and charged 
($D^+ {D}^{*-}$) channels independently, which are related to the
isoscalar and isovector channels by
\begin{eqnarray}
| D\bar{D^*}\,[0] \rangle = \frac{1}{\sqrt{2}}\,\left[
| D^{0}\bar{D}^{*0} \rangle  + | D^+ {D}^{*-} \rangle \right] \, , \\
| D\bar{D^*}\,[1] \rangle = \frac{1}{\sqrt{2}}\,\left[
| D^0\bar{D}^{0*} \rangle  - | D^+ {D}^{*-} \rangle \right] \, ,
\end{eqnarray}
and assume that isospin symmetry is only broken at the level of the masses.
If we consider the basis of physical states
\begin{eqnarray}
\mathcal{B}_{X(3872)} = \left\{ 
\frac{1}{\sqrt2}\left (| D^0\bar{D}^{*0} \rangle-| D^{*0}\bar{D}^{0}
\rangle\right)\right. ,\nonumber \\ 
\left. \frac{1}{\sqrt2}
\left ( | D^+ D^{*-} \rangle-| D^{*+} {D}^{-} \rangle\right)  \right\} \, ,
\end{eqnarray}
we find that we can express the EFT potential as
\begin{eqnarray}
\label{eq:VX-LO}
V^{\rm LO} = \frac{1}{2}
\begin{pmatrix}
C_0 + C_1 & C_0 - C_1 \\
C_0 - C_1 & C_0 + C_1
\end{pmatrix} \, ,
\end{eqnarray}
where $C_0$ and $C_1$ stand for
\begin{eqnarray}
\label{eq:CX-LO}
C_0 &=& C_{0a} + C_{0b} \,,  \qquad C_1 = C_{1a} + C_{1b} \, , 
\end{eqnarray}
that is, the linear combination of isoscalar and isovector counterterms
corresponding to the $1^{++}$ channel.

\section{Isospin Symmetry Violation in the $X(3872)$}
\label{sec:isobreaking}

The Belle collaboration reported for the first time the
decays of the $X(3872)$ into the (isoscalar) $J / \Psi \pi^+ \pi^- \pi^0$
and the (isovector) $J / \Psi \, \pi^+ \pi^-$
channels~\cite{Abe:2005iya}. The non-negligible size of the latter one
hints to the existence of  isospin breaking terms in the dynamics that
govern these decays.
The size of the violation is remarkable, as indicated by
the branching ratio~\cite{Choi:2011fc}
\begin{eqnarray}
\mathcal{B}_{X} = \frac{\Gamma (X(3872) \to J / \Psi \, \pi^+ \pi^- \pi^0)}
{\Gamma (X(3872) \to J / \Psi \, \pi^+ \pi^-)} = 0.8 \pm 0.3 \, , \nonumber \\
\label{eq:Bx-ratio}
\end{eqnarray}
where the central value is even smaller than one.
The most natural explanation for the large ratio is the isospin breaking
generated by the mass difference of the neutral ($D^0 \bar{D}^{0*}$)
and charged ($D^+ D^{*-}$) channels in the
$X(3872)$~\cite{Gamermann:2009fv,Gamermann:2009uq}, which would not
have  a definite isospin. In this picture, at short $D\bar D^*$ distances, the
$X(3872)$ would be a linear combination of $I=0$ and $I=1$
components. The $J / \Psi\, 3\pi$ and $J / \Psi\, 2\pi$ decays would proceed
through the isospin invariant coupling of these final  states to the appropriated short
distance $X(3872)$ isospin wave function components~\cite{Gamermann:2009uq}.

We expect the $2\pi$ and $3\pi$ decays to happen via an intermediate
$\rho$ and $\omega$ meson.
In this regard it is interesting to notice the careful analysis of
Hanhart et al.~\cite{Hanhart:2011tn}, in which the branching ratio
$\mathcal{B}_X$ is translated into the more convenient ratio 
\begin{eqnarray}
R_X =
\frac{\mathcal{M}(X \to J / \Psi \,\rho)}{\mathcal{M}(X \to J / \Psi \, \omega)}
= 0.26^{+0.08}_{-0.05} \, , \label{eq:ratio}
\end{eqnarray}
that, instead of the decay widths, involves the amplitude of the $X(3872)$
to decay into $J / \Psi$ and the $\rho$ or $\omega$ mesons.
We stress that the amplitude ratio $R_X$ is equivalent to the experimental
branching ratio $\mathcal{B}_X$, provided we assume the quantum numbers
of the  $X(3872)$ to be $J^{PC} = 1^{++}$
(see Ref.~\cite{Hanhart:2011tn} for further details).
In particular the experimental errors in $\mathcal{B}_X$ are completely
accounted for by the errors in $R_X$.
In Ref.~\cite{Hanhart:2011tn}, the transition amplitudes for the decays
of the $X(3872)$ into $J / \Psi$ and a vector meson are parametrized as
\begin{eqnarray}
\mathcal{M}(X \to J / \Psi \, V) = g_{X(V)} \, f_{X}(p) ,
\end{eqnarray}
where $V = \omega, \rho$ and $f_{X}$ is a Blatt-Weisskopf barrier factor
that depends on the momentum $p$ of the $J / \Psi$ in the $X$ rest frame.
For the $1^{++}$ assignment, we have $f_{X}(p) = 1$ yielding the value quoted
in Eq.~(\ref{eq:ratio}) for $R_X$.
Other $J^{PC}$ assignments of the $X(3872)$, in particular $J^{PC} = 2^{-+}$,
will imply a different form for the decay amplitude and consequently 
a change in the value of $R_X$~\cite{Hanhart:2011tn}.

If we assume that the $X(3872)$ is a $D\bar{D}^*$ molecule with quantum
numbers $1^{++}$, we can calculate $g_{X(\omega)}$ and $g_{X(\rho)}$ 
in terms of the analogous $g_{\omega}$ and $g_{\rho}$ couplings
for the free $D\bar{D}^*$ meson pair, which are given by
\begin{eqnarray}
\mathcal{M}(D\bar{D}^* (IS) \to J / \Psi \, \omega) &=& g_{\omega} \, , \\
\mathcal{M}(D\bar{D}^* (IV) \to J / \Psi \, \rho) &=& g_{\rho} \, ,
\end{eqnarray}
where $IS$ and $IV$ indicate the isoscalar and isovector configurations
respectively.
When the mesons are bound in the $X(3872)$, we can define the isoscalar
and isovector wave functions as the brackets
\begin{eqnarray}
\langle X | D\bar{D}^*(IS)\, \vec{q}\,\rangle &=& {\Psi}_{X(IS)}(\vec{q}\,) \, ,
\\
\langle X | D\bar{D}^*(IV)\, \vec{q}\,\rangle &=& {\Psi}_{X(IV)}(\vec{q}\,) \, , 
\end{eqnarray}
where we have explicitly indicated the relative momentum $\vec{q}$
of the meson-antimeson pair.
Now, we can relate the $g_{\omega}$ and $g_{\rho}$ couplings with the 
$g_{X(\omega)}$ and $g_{X(\rho)}$ ones by noticing that we can rewrite
the decay amplitude $\mathcal{M}(A \to B)$ as the bracket
$\langle B | A \rangle$.
Then, we insert the two-body identities
\begin{eqnarray}
{\bf 1}_{IS} &=& \int \frac{d^3 \, \vec{q}}{(2\pi)^3}\,
| D\bar{D}^*(IS)\, \vec{q}\, \rangle \, \langle D\bar{D}^*(IS)\, \vec{q}\, | \, , \\
{\bf 1}_{IV} &=& \int \frac{d^3 \, \vec{q}}{(2\pi)^3}\,
| D\bar{D}^*(IV)\, \vec{q}\, \rangle \, \langle D\bar{D}^*(IV)\, \vec{q}\, | \, ,
\end{eqnarray}
for the isoscalar and isovector channels within the brackets and
end up with
\begin{eqnarray}
\mathcal{M}(X \to J / \Psi \, \omega) &\equiv& g_{X(\omega)} = g_{\omega}\,\hat{\Psi}_{X(IS)} \, , \\ 
\mathcal{M}(X \to J / \Psi \, \rho) &\equiv& g_{X(\rho)} = g_{\rho}\,\hat{\Psi}_{X(IV)} \, ,
\end{eqnarray}
where $\hat{\Psi}_{X(IS)}$ and $\hat{\Psi}_{X(IV)}$ are defined as
\begin{eqnarray}
\hat{\Psi}_{X(IS)} &=& \int \frac{d^3 \, \vec{q}}{(2\pi)^3}
\Psi_{X(IS)}(\vec{q}\,) \, , 
\\
\hat{\Psi}_{X(IV)} &=& \int \frac{d^3 \, \vec{q}}{(2\pi)^3}
\Psi_{X(IV)}(\vec{q}\,) \, ,
\end{eqnarray}
that is, they are the coordinate space wave functions at the origin
(more properly, around the origin, as the regulator smears
the wave functions)~\footnote{
Alternatively, we can employ the coupling versus wave-function
language of Ref.~\cite{Gamermann:2009uq} to arrive at the same result.}.
Putting all the pieces together, we find that the ratio $R_X$
can be expressed as
\begin{eqnarray}
R_X = \frac{g_{X(\rho)}}{g_{X(\omega)}} =
\frac{g_{\rho}}{g_{\omega}}\,\frac{\hat{\Psi}_{X(IV)}}{\hat{\Psi}_{X(IS)}} \, ,
\end{eqnarray}
However, we are not working in the isospin basis,
but rather in the particle basis.
If we denote the neutral and charged channels by the subscripts 0 and 1,
we have that
\begin{eqnarray} 
\hat{\Psi}_{X0} &=& \frac{1}{\sqrt{2}}\,
\left( \hat{\Psi}_{X(IS)} + \hat{\Psi}_{X(IV)} \right) \, , \\
\hat{\Psi}_{X1} &=& \frac{1}{\sqrt{2}}\,
\left( \hat{\Psi}_{X(IS)} - \hat{\Psi}_{X(IV)} \right) \, ,
\end{eqnarray}
from which we can rewrite $R_X$ as
\begin{eqnarray}
R_X = \frac{g_{\rho}}{g_{\omega}}\,
\frac{\hat{\Psi}_{X0} - \hat{\Psi}_{X1}}{\hat{\Psi}_{X0} + \hat{\Psi}_{X1}} \, .
\end{eqnarray}
At this point we notice that if we know $R_X$, the binding energy of
the $X(3872)$ and the $g_{\rho}/g_{\omega}$ ratio, we can determine
the contact range potential that binds the $X(3872)$ (which in turn
determines whether the $X(3872)$ has an isovector partner).
Of these quantities the only unknown is the $g_{\rho}/g_{\omega}$ ratio,
which can be determined from the SU(3) relation
\begin{eqnarray}
g_{\rho} - g_{\omega} = -\sqrt{2}\,g_{\phi} \, ,
\end{eqnarray}
and the OZI rule, that implies
\begin{eqnarray}
g_{\rho}, g_{\omega} \gg g_{\phi} \, ,
\end{eqnarray}
which amounts to ignoring the decay of the $X(3872)$ into $J/\Psi \phi$.
Thus, we can conclude that $g_{\rho}/g_{\omega} \simeq 1$ and that the
$R_X$ ratio simplifies to~\footnote{
Of course, there are deviations from the result above coming
from the approximate nature of the OZI rule and to a lesser
extend to SU(3) breaking effects.
However, estimating the size of these deviations is not straightforward
and therefore we have not attempted to systematically include
this error source in the calculations to come.
In this regard we simply note that owing to the large relative error
in $R_X$ ($20$-$30\%$) smaller relative deviations from
$g_{\rho}/g_{\omega} \simeq 1$ (e.g. $10$-$20\%$), for which our results turn out
to be little sensitive,  will be
inconsequential as the two errors are independent and thus are added
quadratically.  }
\begin{eqnarray}
R_X =
\frac{\hat{\Psi}_{X0} - \hat{\Psi}_{X1}}{\hat{\Psi}_{X0} + \hat{\Psi}_{X1}} \, .
\label{eq:RX_OZI}
\end{eqnarray}

The next step is to solve the bound state equation for the $X(3872)$.
We consider that the $X(3872)$ is a $D\bar{D}^*$ molecule with quantum numbers
$J^{PC} = 1^{++}$, where we can distinguish between the neutral
($D^0\bar{D}^{*0}$) and charged ($D^{+} D^{*-}$) components
of the wave function.
That is, there are two channels in the bound state equation.
We regularize the ${\rm LO}$ potential of Eq.~(\ref{eq:VX-LO}) with
a Gaussian regulator function and take a cut-off
$\Lambda = 0.5-1.0\,{\rm GeV}$.
The couplings $C_0$ and $C_1$ (or equivalently $C_{0a} + C_{0b}$ and
$C_{1a} + C_{1b}$) are determined by the condition of reproducing
the location of the $X(3872)$ state and the $\rho / \omega$ ratio
$R_X$, from which we obtain the values
\begin{eqnarray}
C_0 = -1.693^{+0.036}_{-0.023}\,{\rm fm}^2 &\qquad& (-0.731^{+0.008}_{-0.005}\,{\rm fm}^2)  \, , \\
C_1 = -0.08^{+0.42}_{-0.41}\,{\rm fm}^2 & \qquad & 
(-0.373^{+0.089}_{-0.090}\,{\rm fm}^2 ) \, , 
\end{eqnarray}
for $\Lambda = 0.5\,{\rm GeV}$ ($1\,\rm GeV$).
The couplings indicate that the strength of the interaction in the isovector
channel is weaker than in the isoscalar one.
In particular, the isovector coupling $C_1$ is not strong enough to generate
a second bound state, the (mostly) isovector partner of the $X(3872)$.

We find it worth commenting the comparison of our results for the counterterms
with the hidden gauge model of Gamermann et
al.~\cite{Gamermann:2009fv,Gamermann:2009uq}, 
in which the value of the counterterms are determined
from meson exchange saturation.
While the hidden gauge model predicts $C_1 = 0$ (as the contribution
from the $\rho$ and $\omega$ mesons cancel out) and $R_X = 0.136$
(for a sharp cut-off regulator and $\Lambda = 653\,{\rm MeV}$,
see Ref.~\cite{Gamermann:2009uq} for details\footnote{Note that, when
  $C_1=0$, once the regulator/renormalization procedure and the cut-off
  is fixed, $R_X$ is totally determined by the mass of the $X(3872)$.} ),
we obtain a small (yet important)
contribution to $C_1$.
As can be seen, this small contribution is necessary to fine tune
the isospin violating branching ratio of the $X(3872)$ decays
to its exact value.

\section{The SU(3) and HQSS Partners of the $X(3872)$}
\label{sec:partners}

If we are able to determine the value of the counterterms of
the ${\rm LO}$ EFT,
we can calculate the location of the molecular partners of the $X(3872)$.
There are four parameters.
Isoscalar molecules ($I=0$, $S=0$) without hidden strangeness can be described
with $C_{0a}$ and $C_{0b}$.
Isospinor ($I=1/2$, $S= \pm 1$) and isovector ($I = 1$, $S = 0$) states
are in turn determined by $C_{1a}$ and $C_{1b}$.
Finally, for molecular states with hidden strangeness, the contact range
interactions are the average of the isoscalar and isovector ones.

We fix two of the counterterms from the location of the $X(3872)$ resonance and
its isospin breaking branching ratio, as explained in the previous section.
The remaining two require the identification of two partners of the $X(3872)$.
We have chosen the $X(3915)$~\cite{Uehara:2009tx} as a $0^{++}$ isoscalar
$D^*\bar{D}^*$ molecule and the $Y(4140)$~\cite{Aaltonen:2009tz}
as a $0^{++}$ $D_s^*\bar{D}_s^*$ molecule, guided by its apparently
dominant decay into $J / \Psi \phi$. 
We notice that these identifications were proposed for the first time
in Refs.~\cite{Liu:2009ei,Branz:2009yt,Ding:2009vd}.

The $0^{++}$ assignment for the quantum numbers of
the $X(3915)$ and $Y(4140)$ deserve some discussion.
On the one hand, the $2^{++}$ option is excluded:
HQSS fixes the location of the isoscalar $2^{++}$ $D^* \bar{D}^*$
partner of the $X(3872)$ in the vicinity of $4012\,{\rm MeV}$,
far away from the $3915\,{\rm MeV}$ region.
If we additionally consider the isospin breaking decays of the $X(3872)$,
we can determine that the isovector $2^{++}$ $D^* \bar{D}^*$ and
the isoscalar $2^{++}$ $D_s^* \bar{D}_s^*$ molecules
do not exist.
On the other the $1^{+-}$ option can also be discarded
from the decays of the $X(3915)$
and $Y(4140)$ into $J / \Psi \, \omega$ and $J / \Psi \, \phi$,
requiring a positive C-parity state.
Thus we are only left with $J^{PC} = 0^{++}$.

However, the choice of the $X(3915)$ and $Y(4140)$ states as input
is not entirely free of problems.
The first is the binding energy of the $X(3915)$,
approximately $100\,{\rm MeV}$.
Around this binding there is power counting transition
-- one pion exchange changes from perturbative to non-perturbative --
though the critical binding energy that marks this transition is not known
exactly and can happen at larger bindings
than expected~\cite{Valderrama:2012jv}.
In this regard the investigations of Ref.~\cite{Nieves:2012tt}
shows (by performing the explicit calculations) that we are still in the
perturbative regime for the $X(3915)$ and we can employ a
contact theory to describe it at ${\rm LO}$.
The $Y(4140)$, with a binding energy of $80\,{\rm MeV}$, is not affected
by this issue because the $D_s^* \bar{D}_s^*$ system cannot
exchange a single pion.
The second problem is the debatable experimental status of the $Y(4140)$.
After its first and only observation by the CDF
Collaboration~\cite{Aaltonen:2009tz},
subsequent experiments have failed to find it~\cite{Shen:2009vs,Aaij:2012pz}.
This means that the consequences derived from the assumption that the
$Y(4140)$ exists and that it has a molecular structure should be taken
with a grain of salt.

Now we fix the counterterms to the location of
the $X(3915)$ and $Y(4140)$.
For convenience we define first the linear combinations of counterterms:
\begin{eqnarray}
C_2 &=& C_{0a} - 2 C_{0b} \, , \\
C_3 &=& \frac{1}{2}\,\left( (C_{0a} - 2 C_{0b}) + (C_{1a} - 2 C_{1b})
\right)
\, ,\label{eq:c3}
\end{eqnarray}
where $C_2$ / $C_3$ is a convenient way to write the ${\rm LO}$ potential
of the $X(3915)$ / $Y(4140)$ channel.
We obtain
\begin{eqnarray}
C_2 = -6.710\,\,{\rm fm}^2 &\qquad&  (-1.611 \,{\rm fm}^2)\, , \\
C_3 = -5.915\,\,{\rm fm}^2 &\qquad& (-1.459 \,{\rm fm}^2) \, , \label{eq:c3num}
\end{eqnarray}
for a gaussian regulator $\Lambda = 0.5\,{\rm GeV}$ ($1\,\rm GeV$).
We can transform the values of the combinations above (and the corresponding
ones for $C_0$ and $C_1$)
to the standard counterterm representation, yielding
\begin{eqnarray}
C_{0a} = -3.366^{+0.024}_{-0.015}\,{\rm fm}^2 
 &\qquad& (-1.024^{+0.005}_{-0.003} \,{\rm fm}^2)  \, , \label{eq:c0abis}\\
C_{0b} = +1.673^{+0.012}_{-0.008} \,{\rm fm}^2 &\qquad& (+0.293^{+0.004}_{-0.002}\,{\rm fm}^2) \, , \\
C_{1a} = -1.76^{+0.29}_{-0.29} \,{\rm fm}^2 &\qquad& (-0.684^{+0.064}_{-0.063} \,{\rm fm}^2) \, , \\
C_{1b} = +1.68^{+0.15}_{-0.15} \,{\rm fm}^2 &\qquad& (+0.311^{+0.033}_{-0.033} \,{\rm fm}^2 )\, , \label{eq:c1b}
\end{eqnarray} 
for $\Lambda = 0.5\,{\rm GeV}$ ($1\,\rm GeV$), where the error comes
from the uncertainty in $R_X$ (a negligible effect in the isoscalar
channels, but important in the isovector ones).

We notice that there are two additional error sources:
the violations of HQSS due to the finite charm quark mass
and the breaking of light flavour symmetry.
The first of these effects can be taken into account by noticing
that the EFT potential has a relative uncertainty of the order of
\begin{eqnarray}
V^{\rm LO}_{(m_Q = m_c)} = V^{\rm LO}_{(m_Q \to \infty)}\,
(1 \pm \frac{\Lambda_{QCD}}{m_c}) \, ,
\end{eqnarray}
with respect to the exact heavy quark limit.
Taking a value of around  1.5 GeV for the charm quark mass  and
$\Lambda_{\rm QCD} \sim 200$ MeV, we should expect a
15\% violation of HQSS for the LO contact range
potentials~\footnote{Actually, the 15\% violation represents the full
expected deviation from the heavy quark limit in the charm sector.
Even though heavy quark symmetry involves heavy flavour symmetry as well
as HQSS, only the later one is relevant for our purposes.
Thus, when we are talking about HQSS violations, it is merely
language abuse for heavy quark symmetry violations in general.
}.
 
The second error source affects molecules containing strange
quarks~\footnote{We do not consider isospin breaking effects for the
  potential, as their size will be negligible.}, where we expect the
contact range potential to deviate slightly from the pure SU(3)
prediction, that is
\begin{eqnarray}
V^{\rm LO}_{(s-quark)} = V^{\rm LO}_{SU(3)}\,
(1 \pm \delta_{SU(3)}) \, .
\end{eqnarray}
In the expression above $\delta_{SU(3)}$ is the relative size of the
SU(3)-breaking effects, which can be estimated from the ratio of the
kaon and pion decay constants $f_K / f_{\pi} \sim 1.2$, yielding
$\delta_{SU(3)} = 0.2$.
This uncertainty also affects the determination of the $C_3$ counterterm
from the mass of the $Y(4140)$ state.
Thus we can assume a 20\% relative error in the value we give
for this parameter in Eq.~(\ref{eq:c3num}).
In turn, this will translate into an additional error
in the isovector counterterms $C_{1a}$ and $C_{1b}$,
see Eq.~(\ref{eq:c3}) for details.
Actually, their errors are fully anticorrelated to ensure that the linear
combination $C_1 = C_{1a} + C_{1b}$ is free from the SU(3) uncertainties
of $C_3$.
For simplicity and due to the exploratory nature of the present work
we have neglected these correlations.
Instead we have subtituted the correlated error in $C_{1a}$ and $C_{1b}$
by an uncorrelated error in $C_{1a}$~\footnote{Note that we can rewrite $C_3$
as $\frac{1}{2} C_2 - C_1 + \frac{3}{2}\,C_{1a}$.
The isoscalar part of the interaction $C_2$ and the isovector $C_1$
are fixed by the $X(3872)$ and $X(3915)$ inputs, meaning that
all the error can be transferred to $C_{1a}$.
As a matter of fact the decomposition is not unique: we could have transferred
the error to $C_{1b}$ instead.
},
a choice that overestimates the size of
the errors in the molecular masses.

At this point we find it worth mentioning that the three error sources
we have considered -- the experimental error in $R_X$ plus the breaking
of HQSS and $SU(3)$ light flavour symmetry -- are independent:
we can compute the total error by adding the partial errors in quadratures.

Finally, we emphasize that the choice of a regulator is inessential
in EFT calculations.
It does not matter which regulator we have chosen,
as far as the cut-off window is sensible enough.
We have explicitly tested this assumption by calculating the full molecular
spectrum with other regulators.
In general we find small changes in the central location of the states
that are compatible with the cut-off uncertainty we already
find in Tables~\ref{tab:hqs-partners-isoscalar}-\ref{tab:hqs-partners-strange},
i.e. around $10\,{\rm MeV}$ for the most tightly bound cases.
For instance, the location of the $1^{+-}$ isoscalar molecule we predict
in Table~\ref{tab:hqs-partners-isoscalar} changes
from $3955\,{\rm MeV}$ ($3958\,{\rm MeV}$) for a gaussian regulator
to $3954\,{\rm MeV}$ ($3957\,{\rm MeV}$) with a sharp cut-off,
$3965\,{\rm MeV}$($3964\,{\rm MeV}$) with a monopolar regulator
and $3960\,{\rm MeV}$($3961\,{\rm MeV}$) with a dipolar regulator,
in all cases with a cut-off $\Lambda = 0.5\,{\rm GeV}$
($1.0\,{\rm GeV}$)~\footnote{Actually, one should take
into account the optimal cut-off window depends on the choice of
a regulator. In particular, for the monopolar regulator we should
use larger cut-offs than for the sharp cut-off and gaussian cases.
However, for the purposes of the current discussion we can
ignore this effect and take the same cut-off value for all
the regulators we have considered.}.

\subsection{The SU(2) Isoscalar ($I=0$) Partners}
\label{sec:isos}

We begin with the SU(2) isoscalar sector,
in which we ignore the hidden strange components.
The states are determined by the counterterms
$C_{0a}$ and $C_{0b}$.
We do not take into account particle coupled channel effects 
as they are subleading, as explicitly checked in Ref.~\cite{Nieves:2012tt}.
There is one exception though, the $1^{++}$ and $2^{++}$ channels,
where the mass gap between the neutral and charged channels
($8$ and $6\,{\rm MeV}$ in each case)
is similar in size to the binding energy
in the isospin symmetric limit
($4$ and $5\,{\rm MeV}$).
This suggests that we may treat the neutral and charged channels
as explicit degrees of freedom of the theory.
We note that the inclusion of isospin violation is the only difference
with respect to the previous analysis of Ref.~\cite{Nieves:2012tt}.
The spectrum of molecular states is presented in
Table~\ref{tab:hqs-partners-isoscalar}.
As can be seen, isospin violation is a small perturbation over the
former predictions of Ref.~\cite{Nieves:2012tt} (though it is
still crucial to describe isospin violating decays properly).

We, however, would like to make a few remarks. 
The first one concerns the $2^{++}$ state. 
The central values of the counterterms
as given in Eqs.~(\ref{eq:c0abis})--(\ref{eq:c1b})
predict that the $2^{++}$ state lies very close (less than $1$ MeV)
to the lowest energy threshold, i.e. $D^{*0}\bar D^{*0}$. 
When we decrease the strength of the potential to account
for the uncertainties of our approach, the pole reaches the neutral
threshold and then bounces back into the second Riemann sheet.
That is, the state becomes virtual (instead of bound). In any case,
the existence of the pole  will strongly influence the amplitude at threshold.

The second comment is about the $0^{++}$ $D\bar D$ channel.
As can be seen in the Table~\ref{tab:hqs-partners-isoscalar},
this state is bound by about 20-25 MeV. 
For simplicity, we have used in this channel the isospin symmetric limit.
Yet the $D^{0}\bar D^{0}-D^{+} D^{-}$ threshold gap is around 9 MeV,
and it might make sense the explicit consideration of
isospin breaking in this channel.
However, as in the $X(3872)$ and $X(4012)$ cases, the effect is rather small,
justifying the validity of the isospin symmetric limit
for the spectroscopy problem.

Finally, there is a remark concerning the possible effect of hidden
strange channels in the dynamics of the $X(3915)$ state.
As we will see, there is a $0^{++}$ $D_s \bar D_s$ state at $3925\,{\rm MeV}$
(see Table~\ref{tab:hqs-partners-strange}) that in principle can mix
with the $X(3915)$.
However, explicit calculations show that the influence of the hidden strange
channel is numerically marginal: it moves the position of the predicted
molecular states by a small fraction of a MeV
(usually $\Delta M \sim 0.1-0.2\,{\rm MeV}$),
a tiny effect compared to other error sources.
The reason lies in the transition potential between the isoscalar $0^{++}$
$D^* \bar{D}^*$ and $D_s \bar D_s$ channels:
\begin{eqnarray}
\langle D^* \bar{D}^* | V^{\rm LO}(0^{++}) | D_s \bar D_s \rangle =
\sqrt{\frac{3}{2}} (C_{0b}-C_{1b}) \, ,
\end{eqnarray}
which turns out to be numerically quite small, since we accidentally find
$C_{0b} \simeq C_{1b}$, see Eqs.~(\ref{eq:c0abis})--(\ref{eq:c1b}).

\begin{table*}
\begin{center}
\begin{tabular}{|c|c|c|c|c|c|c||c|}
\hline \hline
$J^{PC}$ & $\rm H\bar{H}$ & $^{2S+1}L_J$  & $V_C$
& $E$ $(\Lambda =0.5$ GeV) & $E$ $(\Lambda =$ 1 GeV) & ${\rm
  Exp}$~\cite{Beringer:1900zz} & Threshold [MeV]\\
\hline
$0^{++}$ & $ D\bar{D}$ & $^1S_0$ & $C_{0a}$
& $3709^{+9}_{-10}$ & $3715^{+12}_{-15} $  & $-$ & 3734.5$^*$\\
\hline
$1^{++}$ & $ D^*\bar{D}$ & $^3S_1$ & Eqs.~(\ref{eq:VX-LO}) and (\ref{eq:CX-LO})
& Input & Input & $3871.6$ & 3871.8/3879.9\\
$1^{+-}$ & $ D^*\bar{D}$ & $^3S_1$ & $C_{0a} - C_{0b}$
& $3815 ^{+16}_{-17}$ & $3821^{+23}_{-26}$ & $-$ & 3875.9$^*$\\
\hline
$0^{++}$ & $ D^*\bar{D}^*$ & $^1S_0$ & $C_{0a} - 2\,C_{0b}$ & Input
& Input & $3917$ & 4017.3$^*$ \\
$1^{+-}$ & $ D^*\bar{D}^*$ & $^3S_1$ & $C_{0a} - C_{0b}$
& $3955^{+16}_{-17}$ & $3958^{+24}_{-27}$ & $3942$ & 4017.3$^*$  \\
$2^{++}$ & $ D^*\bar{D}^*$ & $^5S_2$ & Eqs.~(\ref{eq:VX-LO}) and (\ref{eq:CX-LO})
& $4013^{\dagger\dagger}_{-9} $ & $4013^{\dagger\dagger}_{-12} $ & $-$ & 4014.0/4020.6\\
\hline \hline
\end{tabular}
\end{center}
\caption{
Predicted masses (in MeV) of the SU(2) isoscalar  HQSS partners
of the $X(3872)$ resonance for two different values of the Gaussian
cutoff. We use as input 3871.6 MeV, 3917.4 MeV and 4140 MeV for the $X(3872)$,
$X(3915)$ and $Y(4140)$, respectively.  In the last column, we also
give the thresholds for the different charge combination channels
($P^0 \bar P^0 / P^+P^-$)  or the threshold value that results 
when we neglect isospin breaking effects and we 
employ charge averaged masses. In this latter case, we mark it with an asterisk (*).
Errors in the predicted masses are obtained by adding in quadratures 
the uncertainties stemming  from the two sources of systematic errors
discussed at the end in Subsect.~\ref{sec:partners}: errors in our
determination of the LO counterterms in
Eqs.~(\ref{eq:c0abis})--(\ref{eq:c1b}) and variations obtained by modifying the strength of
the contact interaction in each channel by $\pm 15\%$,
which corresponds to the expected violations of HQSS
for the charm quark mass. $\dagger\dagger:$ the upper error of the $2^{++}$
state mass deserves a detailed discussion that can be found in the
text (Subsect.~\ref{sec:isos}).} 
\label{tab:hqs-partners-isoscalar}
\end{table*}

\subsection{The Isospinor ($I=\frac{1}{2}$) Partners}
\label{sec:i12}
The isospinor molecules are different in the sense that they have not
well-defined C-parity, as they are not bound states of a heavy meson
and its antimeson.
In general this poses no problem (the formalism is identical to the one
in the previous case) except for the $1^{+}$ $D_s \bar D^*$ and $D
\bar D_s^*$ molecules.
The $D_s \bar D^*$ and $D \bar D^*_s$ thresholds are separated by only $2\,{\rm MeV}$
and require a coupled channel treatment.
The ${\rm LO}$ potential reads
\begin{eqnarray}
\label{eq:VISO-LO}
V^{\rm LO} = 
\begin{pmatrix}
\phantom{+}C_{1a} & -C_{1b} \\
-C_{1b} & \phantom{+}C_{1a}
\end{pmatrix} \, ,
\end{eqnarray}
where channel $1$ ($2$) corresponds to $D_s \bar D^*$ ($D_s^* \bar D$).
We find that this coupled channel potential only leads to one bound state
(of a maximum of two).
We list this and the other isospinor molecules
in Table \ref{tab:hqs-partners-isospinor}, where we have considered
only the strangeness one states. The spectrum is identical for the
strangeness minus one sector. 
We also notice that the error bands are bigger
as they include the additional SU(3) breaking
effects at the level of the contact range interaction.

There is a total of four states, one of them -- the $0^+$ $D_s
\bar{D}$ molecule -- almost at threshold.
Actually, there exists a small violation of the third component of isospin,
because of the different masses of the $D^-$ ($D^{*-})$ and $\bar D^0$
($\bar D^{*0}$) mesons.
In general the isospin splitting is smaller than the other errors
in the calculations and we ignore it.
The exception is the $0^+$ $D_s\bar{D}$ molecule:
in Table \ref{tab:hqs-partners-isospinor}
we only report the $I_3=-1/2$ component of the molecule,
which is being formed by the $D_s^+  D^-$ interaction~\footnote{The dynamics of
  the other state member of the isospin doublet is similar, being its
  mass just shifted by about 5 MeV. This shift is due to the mass
  difference between the  $D_s^+ D^-$ and the $D_s^+  \bar
  D^0$ pairs. }. 
Similarly to what happens with the isoscalar $2^{++}$ state,
with the central values of the counterterms in Eq.~(\ref{eq:c0abis})--(\ref{eq:c1b})
we predict a $0^+$ state that lies very close to
threshold. When we decrease the strength of the potential, the
molecule approaches the threshold and finally becomes a virtual state,
at least within our simple scheme~\footnote{
A resonance state in one channel is usually
associated to a barrier in coordinate space which is not reproduced by
a constant (in energy) potential. The situation may be different when we
have coupled channels~\cite{YamagataSekihara:2010pj}, in which case one
of the channels can decay into the other.}.  In the
real world, it might well happen that in this situation the state
could become a narrow resonance, placed very close
to threshold, and decaying into the $D_s^+ \bar D^-$ pair instead of a
becoming a virtual state.
In any case, we predict the existence of some structure close
to threshold. The upper errors for the mass of this 
state, as quoted in Table \ref{tab:hqs-partners-isospinor},
just account for the distance between its central mass value
and the $D_s^+ \bar D^-$ threshold.

\begin{table*}
\begin{center}
\begin{tabular}{|c|c|c|c|c|c|c||c|}
\hline \hline
$J^{P}$ & $\rm H\bar{H}$ & $^{2S+1}L_J$  & $V_C$
& $E$ $(\Lambda =0.5$ GeV) & $E$ $(\Lambda =$ 1 GeV) & ${\rm
  Exp}$~\cite{Beringer:1900zz} & Threshold [MeV]\\
\hline
$0^{+}$ & $ D_s^+\bar D^-$ & $^1S_0$ & $C_{1a}$
& $3835.8^{+2.3}_{-7.3}$ & $3837.7^{+0.4}_{-8.1} $  & $-$ & 3838.1\\
\hline
$1^{+}$ & $D_s \bar{D}^*$,$ D_s^*\bar{D}$ & $^3S_1$ & Eq.~(\ref{eq:VISO-LO})
& $3949^{+20}_{-21}$ & $3957^{+22}_{-32}$ & $-$ & $3977.15^\dagger, 3979.55^\dagger$ \\
\hline
$0^{+}$ & $ D_s^*\bar{D}^*$ & $^1S_0$ & $C_{1a} - 2\,C_{1b}$ &
$4056^{+31}_{-35}$
& $4061^{+45}_{-54}$ & $-$ & 4120.9$^\dagger$\\
$1^{+}$ & $ D_s^*\bar{D}^*$ & $^3S_1$ & $C_{1a} - C_{1b}$
& $4091^{+19}_{-22}$ & $4097^{+24}_{-33}$ & $-$ &4120.9$^\dagger$ \\
$2^{+}$ & $ D_s^*\bar{D}^*$ & $^5S_2$ & $C_{1a} + C_{1b}$
& $-$ & $-$ & $-$ & \\
\hline \hline
\end{tabular}
\end{center}
\caption{
Predicted masses (in MeV) of the 
isospinor ($I=\frac{1}{2}$)   HQSS partners of the $X(3872)$ resonance,
for two different values of the Gaussian cutoff. The meaning of the
quoted errors in the table 
is the same as in Table \ref{tab:hqs-partners-isoscalar}. We also give
the relevant thresholds (in MeV) for each channel. We use isospin
third component averaged masses for those cases marked with a
$\dagger$ symbol. Note that as we decrease the  strength of
the potential, the $D_s^+\bar D^-$ state becomes virtual (see
discussion in Subsect.~\ref{sec:i12}).} 
\label{tab:hqs-partners-isospinor}
\end{table*}

\subsection{The Isovector ($I=1$) Partners}

The spectrum of the isovector molecules is similar to the isospinor one.
The reason is that the ${\rm LO}$ potential is identical in both cases,
with the exception of the $1^{++}$ and $2^{++}$ molecules 
owing to isospin violation, as we commented previously.
We find four molecular states of a possible total of six
that we list in Table \ref{tab:hqs-partners-isovector}.
The location of the states is similar to the isoscalar sector, only
that they are a bit less bound.
The two missing states would correspond to the isovector partners of
the $X(3872)$ and $X(4012)$. 

We also mention that there is a complication with the isovector $0^{++}$
$D\bar D$ molecule, which lies very close to the $D\bar D$ threshold.
This implies that explicit isospin breaking should be taken into account.
However, we notice that this is only necessary for the $I_3 = 0$ component
of the isospin triplet, where the neutral ($D^0 \bar{D}^0$) and
charged ($D^{+} D^{-}$) channels are to be found.
If we consider instead the $I_3 = \pm 1$ states (corresponding to the particle
channels $D^{+} \bar{D}^0$ and $D^0 D^{-}$), there is no mixing with the
isoscalar components and no need for coupled channel dynamics.
Thus, for simplicity, we have decided to report only the $I_3=+1$ state
in Table \ref{tab:hqs-partners-isovector}.

\begin{table*}
\begin{center}
\begin{tabular}{|c|c|c|c|c|c|c||c|}
\hline \hline
$J^{PC}$ & $\rm H\bar{H}$ & $^{2S+1}L_J$  & $V_C$
& $E$ $(\Lambda =0.5$ GeV) & $E$ $(\Lambda =$ 1 GeV) & ${\rm
  Exp}$~\cite{Beringer:1900zz} & Threshold [MeV]\\
\hline
$0^{++}$ & $ D^+\bar{D}^0$ & $^1S_0$ & $C_{1a}$
& $3732.5^{+2.0}_{-6.9}$ & $3734.3^{+0.2}_{-6.9} $  & $-$ & 3734.5\\
\hline
$1^{++}$ & $ D^*\bar{D}$ & $^3S_1$ & Eqs.~(\ref{eq:VX-LO}) and (\ref{eq:CX-LO})
& $-$ & $-$ & $-$ &  \\
$1^{+-}$ & $ D^*\bar{D}$ & $^3S_1$ & $C_{1a} - C_{1b}$
& $3848^{+15}_{-17}$ & $3857^{+15}_{-22}$ & $-$ & 3875.9$^*$ \\
\hline
$0^{++}$ & $ D^*\bar{D}^*$ & $^1S_0$ & $C_{1a} - 2\,C_{1b}$ & $3953 ^{+24}_{-26}$
& $3960^{+31}_{-37}$ & $-$ &4017.3$^*$ \\
$1^{+-}$ & $ D^*\bar{D}^*$ & $^3S_1$ & $C_{1a} - C_{1b}$
& $3988^{+15}_{-17}$ & $3995^{+17}_{-23}$ & $-$ &4017.3$^*$ \\
$2^{++}$ & $ D^*\bar{D}^*$ & $^5S_2$ & Eqs.~(\ref{eq:VX-LO}) and (\ref{eq:CX-LO})
& $-$ & $-$ & $-$ &\\
\hline \hline
\end{tabular}
\end{center}
\caption{Predicted masses (in MeV) of the SU(2) isovector  HQSS partners
of the $X(3872)$ resonance for two different values of the Gaussian cutoff. 
The meaning of the quoted errors  in the
table is the same as in Table \ref{tab:hqs-partners-isoscalar}. We also give
the relevant thresholds (in MeV) for each channel. For the $
D^*\bar{D}$ and $ D^*\bar{D}^*$ cases, we give the threshold that
would correspond to the zero isospin third component and calculated
with charge averaged masses. Note that as we decrease the  strength of
the potential, the $D^+\bar D^0$ state becomes virtual. The upper
errors for the mass of this  state just account for the distance
between its central mass value and the threshold. }
\label{tab:hqs-partners-isovector}
\end{table*}

\subsection{The Hidden Strange Partners}

Finally, we consider the molecules with hidden strangeness.
In this sector the strength of the $\rm LO$ potential is the arithmetic
mean of the isoscalar and isovector one.
This means that if there is a bound state in the isoscalar and isovector
sector, we can be confident about the existence of a heavy meson molecule
containing a $s\bar{s}$ quark-antiquark pair.
Conversely, if we consider the $X(3872)$ and $X(4012)$ molecules,
the fact that they have no isovector partners is a strong hint
that there will be no hidden strange partner either.
The four states we obtain are listed in Table \ref{tab:hqs-partners-strange}.

\begin{table*}
\begin{center}
\begin{tabular}{|c|c|c|c|c|c|c||c|}
\hline \hline
$J^{PC}$ & $\rm H\bar{H}$ & $^{2S+1}L_J$  & $V_C$
& $E$ $(\Lambda =0.5$ GeV) & $E$ $(\Lambda =$ 1 GeV) & ${\rm
  Exp}$~\cite{Beringer:1900zz}  & Threshold [MeV]\\
\hline
$0^{++}$ & $ D_s\bar{D}_s$ & $^1S_0$ & $\frac{1}{2}(C_{0a} + C_{1a})$
& $3924^{+10}_{-13}$ & $3928^{+9}_{-19} $  & $-$ & 3937.0 \\
\hline
$1^{++}$ & $ D_s^*\bar{D}_s$ & $^3S_1$ & $\frac{1}{2}(C_{0a} + C_{1a} + C_{0b} + C_{1b})$
& $-$ & $-$ & $-$&  \\
$1^{+-}$ & $ D_s^*\bar{D}_s$ & $^3S_1$ & $\frac{1}{2}(C_{0a} + C_{1a} - C_{0b} - C_{1b})$
& $4035^{+23}_{-25}$ & $4040^{+33}_{-39}$ & $-$ & 4080.8 \\
\hline
$0^{++}$ & $ D_s^*\bar{D}_s^*$ & $^1S_0$ & $\frac{1}{2}(C_{0a} + C_{1a} - 2\,C_{0b} - 2\,C_{1b})$
 & Input
& Input & $4140$  & 4224.6 \\
$1^{+-}$ & $ D_s^*\bar{D}_s^*$ & $^3S_1$ & $\frac{1}{2}(C_{0a} + C_{1a} - C_{0b} - C_{1b})$
& $4177^{+23}_{-25} $ & $4180^{+35}_{-40}$ & $-$ & 4224.6\\
$2^{++}$ & $ D_s^*\bar{D}_s^*$ & $^5S_2$ & $\frac{1}{2}(C_{0a} + C_{1a} + C_{0b} + C_{1b})$
& $-$ & $-$ & $-$ &\\
\hline \hline
\end{tabular}
\end{center}
\caption{
Predicted masses (in MeV) of the  hidden strange  isoscalar HQSS partners
of the $X(3872)$ resonance for two different values of the Gaussian cutoff. 
 The meaning of the
quoted errors is the same as in Table \ref{tab:hqs-partners-isoscalar}. We also give
the relevant thresholds (in MeV) for each channel. }  
\label{tab:hqs-partners-strange}
\end{table*}

\section{Conclusions}
\label{sec:conclusion}

In this work we have shown how the heavy quark spin and light SU(3) flavour
symmetries constrain the charmed meson-antimeson interaction
($H\bar H \to H\bar H$, being $H=D^+,D^0,D_s^0, D^{*+},D^{*0}$ and $D_s^{*0}$).
This has been done within the EFT framework, where the heavy meson interactions
can be easily arranged from more to less relevant thanks to power counting,
the ordering principle behind EFT.
The bottom line of the EFT approach is that contact interactions
(i.e. four meson vertices) dominate the low energy dynamics of
heavy meson molecules.
In turn, SU(3) flavour symmetry and HQSS reduce the number of contact
interactions from twenty four to only four.
The approach we advocate is actually an extension of the ideas
of Ref.~\cite{Nieves:2012tt} for the SU(2) isoscalar channel
to the isospinor, isovector and hidden strange sectors.

The four counterterms can be determined as follows:
we fix three of them by identifying the $X(3915)$, $Y(4140)$ and $X(3872)$
resonances as molecular states.
In particular we consider the $X(3915)$ and $Y(4140)$ to be a $0^{++}$ isoscalar
$D^* \bar D^{*}$ and $D^*_s \bar D^{*}_s$ molecules respectively, while the
$X(3872)$ is a $1^{++}$ isospin admixture $D \bar D^{*}$ state.
Finally, the fourth counterterm is determined from the isospin breaking
branching ratio of the $X(3872)$ into $J / \Psi\,2\pi$
and $J / \Psi\,3 \pi$.
We notice that the $X(3872)$ is weakly bound ($\sim 0.2$ MeV)
and lies extremely close to the $ D^0 \bar{D}^{0*}$ threshold.
Its binding energy is much smaller than the mass splitting
between the charged and neutral components of the $X(3872)$
($\sim 8$ MeV).
Hence, we have taken into account these degrees of freedom explicitly.
We work within a scheme where all interactions are isospin invariant,
but where the $X(3872)$ does not have a well defined
isospin~\footnote{The isospin
mixing depends on the relative distance between the pseudoscalar and
vector mesons.} as a consequence of the mass and kinetic terms
of the $D \bar{D}^{*}$ Hamiltonian~\cite{Gamermann:2009fv,Gamermann:2009uq}.
The isospin breaking of the masses naturally explains the appearance of
an isospin violation (and its quantity) in the decays of the $X(3872)$,
as pointed out in Refs.~\cite{Gamermann:2009fv,Gamermann:2009uq}
for the first time.
However, in the aforementioned works the experimental branching
ratio~\cite{Choi:2011fc}
\begin{eqnarray}
\frac{\Gamma (X(3872) \to J / \Psi \, \pi^+ \pi^- \pi^0)}
{\Gamma (X(3872) \to J / \Psi \, \pi^+ \pi^-)} = 0.8 \pm 0.3 \, , \nonumber 
\end{eqnarray}
is not perfectly reproduced~\footnote{Notice, however,
that Refs.~\cite{Gamermann:2009fv,Gamermann:2009uq} used an older
experimental determination of the branching ratio ($1.0 \pm 0.3 \pm 0.4$)
with much larger errors. It is only with the updated branching ratio
that their explanation fails at the fine quantitative level.}.
The reason is that the authors of Refs.~\cite{Gamermann:2009fv,Gamermann:2009uq}
assume that the isovector $D\bar D^*$ interaction vanishes.
Here, we improve on that and take advantage of the experimental ratio
to constraint the non-vanishing interaction in the isovector channel.

Once we have fixed the counterterms, we have established the existence
and the location of up to a total of 15 molecular partners
of the  $X(3915)$, $Y(4140)$ and $X(3872)$ states, see Tables
\ref{tab:hqs-partners-isoscalar}-\ref{tab:hqs-partners-strange}.
These predictions are subjected to a series of uncertainties,
in particular the approximate nature of HQSS (especially
in the charm sector). 
We have estimated the size of these corrections and concluded that
the HQSS pattern of molecular states is stable, though the exact
location of the states can change by a few tens of ${\rm MeV}$
in certain cases.

Actually, the family of $D^{(*)}\bar D^{(*)}$ states we theorize
depends on the assumption that the $X(3872)$, $X(3915)$
and $Y(4140)$ resonances are molecular. 
While in the $X(3872)$ case the molecular interpretation is compelling
and widely accepted, the $X(3915)$ and $Y(4140)$ states are
merely compatible with it.
Regarding the $X(3915)$, it is interesting to notice that the size of
the decay width of this resonance is difficult to conciliate
with the hypothesis that it is a charmonium~\cite{Guo:2012tv}.
This observation enhances the prospects that the $X(3915)$
may be a molecule after all.
However, with the $Y(4140)$ we have a more serious problem:
this resonance is far from being confirmed experimentally.
Thus not all the states we predict are equally likely.
Predictions derived from the $X(3872)$ should be regarded as more solid
than those depending on the $X(3915)$, which in turn are less
speculative that the ones obtained from the $Y(4140)$.
In this regard, as stressed in Ref.~\cite{Nieves:2012tt}, 
the $2^{++}$ ${\rm D^* \bar{D}^*}$ isoscalar partner of
the $X(3872)$ is still the most reliable prediction
of the present work, followed by the other
isoscalar states.
If in the future we count with clearer molecular candidates
than the $X(3915)$ and, particularly, the $Y(4140)$ resonances,
they could be included in the current scheme instead of the later ones,
helping to achieve more robust predictions.
Conversely, the observation of any of the states predicted here
can serve as proof of the molecular nature of the previous
mentioned resonances.

Finally, we find it interesting to compare our results with those of
the hidden gauge formalism, another theoretical approach for
the study of hidden charm states. 
While the spectrum of the isoscalar molecules in the hidden gauge is similar
to the one we obtain (with the notable exception of the $2^{++}$
state), fewer poles are reported in the $I=1/2$, $I=1$
and hidden strange sectors~\cite{Gamermann:2006nm,Gamermann:2007fi,
Molina:2009ct}. 
If we consider the case of two heavy pseudoscalar mesons,
Gamermann et al.~\cite{Gamermann:2006nm} predict an isoscalar $0^{++}$
$\rm D\bar D$ state in the vicinity of $3700\,{\rm MeV}$.
We can identify this state with the $X(3710)$ $\rm D \bar D$ molecule
we obtain in the present work.
However, they do not predict the existence of the isospinor, isovector
and hidden strangeness partners of the $X(3710)$.
The reason is that in Ref.~\cite{Gamermann:2006nm} the interaction
in the isovector channel is zero.
The same comments apply to the extension of the hidden gauge formalism
to pseudoscalar-vector molecules~\cite{Gamermann:2007fi},
where it is found a counterpart of the $1^{+-}$ $\rm D\bar D^*$
isoscalar molecule that we obtain at $3820\, {\rm MeV}$
at a slightly higher energy ($3840\,{\rm MeV}$)
but no $I=1/2$, $I=1$ or hidden strangeness states.
Last, in the case of two heavy vector mesons the hidden gauge
predicts a series of isoscalar $0^{++}$, $1^{+-}$ and $2^{++}$
$\rm D^* \bar D^*$ states, plus a few non-isoscalar ones~\cite{Molina:2009ct}.
In the isoscalar sector, the $0^{++}$ and $1^{+-}$ $\rm D^* \bar D^+$ resonances
are located in the region around $3943$ and $3945\,{\rm MeV}$ respectively,
not very different to the masses we use ($3917$ and $3955\,{\rm MeV}$).
However, there is a striking difference in the mass of the $2^{++}$
${\rm D^* \bar{D}^*}$ isoscalar state, which in Ref.~\cite{Molina:2009ct}
happens at $3922 \,{\rm MeV}$ (instead of $4013\,{\rm MeV}$).
The reason for this value, which is incompatible with HQSS, is the remarkably
strong vector-vector interaction that is obtained in the hidden gauge model. 
Probably as a result of this strong interaction,
Ref.~\cite{Molina:2009ct} also report the existence of broad
$2^{++}$ isovector $D^*\bar D^*$ and hidden strange $D^*_s\bar D^*_s$
resonances, with widths above 100 MeV, and masses of around 3910 and 4160 MeV,
respectively.  
These states are difficult to accommodate within our HQSS scheme.

\begin{acknowledgments}
MPV thanks the support of the CPAN Programme and the hospitality of the IFIC
during part of this work. CHD thanks the support of the JAE-CSIC
Programme. 
This research was supported by DGI and FEDER funds, under contract
FIS2011-28853-C02-02, and the Spanish Consolider-Ingenio 2010 Programme
CPAN (CSD2007-00042),  by Generalitat Valenciana under contract
PROMETEO/20090090 and by the EU HadronPhysics2 project,
grant agreement no. 227431. 
\end{acknowledgments}


%

\end{document}